\documentclass[a4paper,superscriptaddress,twocolumn,showkeys]{revtex4}

\usepackage[left=1.5cm, right=1.5cm,top=2.5cm,bottom=2cm]{geometry} 
 
\usepackage[utf8]{inputenc}
\usepackage{amsmath}
\usepackage{amsfonts}
\usepackage{amssymb}
\usepackage{graphicx}
\usepackage{commath}

\setcitestyle{numbers}
\usepackage{color}
\usepackage[normalem]{ulem} 

\newcommand{\Replace}[2]{\bgroup\noindent\textcolor{blue}{\xout{#1} #2}\egroup\ignorespacesafterend}
\newcommand{\Delete} [1]{\bgroup\noindent\textcolor{blue}{\xout{#1}}\egroup\ignorespacesafterend}
\newcommand{\Insert} [1]{\bgroup\noindent\textcolor{blue}{#1}\egroup\ignorespacesafterend}
\newcommand{\Comment}[1]{\definecolor{Mygray}{gray}{0.50}\bgroup\color{Mygray}\noindent#1\egroup\ignorespacesafterend}
\newcommand \Stefan [1] {\bgroup\noindent[\textcolor{blue}{\textbf{Stefan}: #1}]\egroup\ignorespacesafterend}
\newcommand \Dominik[1] {\bgroup\noindent[\textcolor{blue}{\textbf{Dominik}: #1}]\egroup\ignorespacesafterend}
\newcommand \Ric[1] {\bgroup\noindent[\textcolor{blue}{\textbf{Riccardo}: #1}]\egroup\ignorespacesafterend}

\newcommand{\DtoC}{\emph{D2C}}
\newcommand{\CDD}{{CDD}}
\newcommand{\hdCDD}{{hdCDD}}

\def\tens#1{\relax\ifmmode\mathsf{#1}\else\textsf{#1}\fi}

\newcommand{\Bb}{{\boldsymbol{\mathnormal b}}}

\newcommand{\Bl}{{\boldsymbol{\mathnormal l}}}
\newcommand{\Bn}{{\boldsymbol{\mathnormal n}}}

\newcommand{\Bq}{{\boldsymbol{\mathnormal q}}}
\newcommand{\Br}{{\boldsymbol{\mathnormal r}}}

\newcommand{\superscr}[1]{\ensuremath{{}^{\rm #1}}}

\newcommand{\Balpha} {\ensuremath{\boldsymbol\alpha}}

\newcommand{\Bxi}    {\ensuremath{\boldsymbol\xi}}

\newcommand{\Bve}    {\ensuremath{\boldsymbol\varepsilon}}

\newcommand{\vp}     {\varphi}

\newcommand{\rhot}   {\ensuremath{\rho\superscr{t}}}
\newcommand{\qt}     {\ensuremath{q\superscr{t}}}

\newcommand{\rhoG}   {\ensuremath{\rho\superscr{G}}}
\newcommand{\Brho}   {\ensuremath{\boldsymbol\rho}}

\newcommand{\RHO}[2] {\ensuremath{\rho^{(#1)}_{#2}}}

\newcommand{\rhoZero}{\ensuremath{\rho^{(0)}}} 
\newcommand{\rhoOne} {\ensuremath{\Brho^{(1)}}} 
\newcommand{\rhoTwo} {\ensuremath{\Brho^{(2)}}}
\newcommand{\qZero}  {\ensuremath{q}^{(0)}} 
\newcommand{\qOne}   {\ensuremath{q}^{(1)}}

\newcommand{\CDDI}   {\ensuremath{\text{CDD}^{(1)}}}
\newcommand{\CDDII}  {\ensuremath{\text{CDD}^{(2)}}}

\newcommand{\CDDn}   {\ensuremath{\text{CDD}^{(n)}}}

\newcommand{\rphi}{{(\Br,\vp)}}        
        
\begin{document}

\title{A Universal Approach Towards Computational Characterization of Dislocation Microstructure}

\author{Dominik Steinberger}
\affiliation{Friedrich-Alexander University of Erlangen-Nuremberg, Department of Materials Science, Dr.-Mack-Str.\ 77, 90762 Fürth, Germany}

\author{Riccardo Gatti}
\affiliation{LEM, UMR 104 CNRS-ONERA, 29 Avenue de la Division Leclerc, 92322 Châtillon, France}

\author{Stefan Sandfeld}
\affiliation{Friedrich-Alexander University of Erlangen-Nuremberg, Department of Materials Science, Dr.-Mack-Str.\ 77, 90762 Fürth, Germany}
\affiliation{corresponding author's email address: stefan.sandfeld@fau.de}

\date{\today}

\begin{abstract}
	Dislocations -- linear defects within the crystal lattice of, e.g., metals -- already have been directly observed and analyzed for nearly a century. While experimental characterization methods can nowadays reconstruct three-dimensional pictures of complex dislocation networks, simulation methods are at the same time more and more able to predict the evolution of such systems in great detail. Concise methods for analyzing and comparing dislocation microstructure, however, are still lagging behind. We introduce a universal microstructure ``language'' which could be used for direct comparisons and detailed analysis of very different experimental and simulation methods.
\end{abstract}

\keywords{Advanced Materials; ICME; Materials Informatics; Data Curation}

\maketitle

\section*{INTRODUCTION}

During most of the 20th century materials science mostly relied on the interplay between theory and experiment in a hypothesis-driven manner. Although numerical tools already contributed in some fields (e.g., atomistic simulations were used from the 1960s on), simulations were not considered as separate discipline due to strongly limited models and computational power.
The last two decades brought a dramatic paradigm shift to materials science with the advent and easy accessibility of powerful computational concepts and refined simulation methods which build the basis of today's computational materials science: while predictive computational models are more and more able to approach time and length scales similar to those observed in experiments, novel data-driven approaches are able to guide experiments, to discover previously unknown mechanisms and to help designing advanced materials. 

To accelerate and direct these efforts of {Integrated Computational Materials Engineering} (ICME) a number of initiatives recently came to life, as, e.g., the {European Materials Modelling Council} (EMMC) and the U.S.\ {Materials Genome Initiative} (MGI), whose goal is ``to decrease the time and cost of the materials discovery to deployment process'' \citep{MGI}. All initiatives emphasize -- in line with demands from fundamental research -- the following key aspects for  design and discovery of new materials:
(i) the requirement of dedicated simulation codes and community-driven software;
(ii) novel integration techniques of experimental and theoretical data throughout research, development, design and manufacturing;
(iii) the faithful bridging of models across length and time scales, e.g., within a multi-scale simulation framework; 
(iv) innovative data-based bridging between different experimental, theoretical and simulation models by a (one-way) information transfer for, e.g., validation purposes or for use as physical initial values from lower scale methods.


Metal plasticity on the micro scale is a typical case where processes on multiple time and length scales interact with each other, e.g., phenomena on small scales may determine properties on the larger device scale: dislocations -- linear defects within the crystal lattice -- are the carrier of plastic deformation and are responsible for a large variety of emergent properties, as, e.g., self-organization of dislocations into complex microstructures, size-dependent mechanical behavior or hardening. Therefore, during the last century significant effort has been dedicated to experimentally observing and characterizing dislocations, as well as to understanding the evolution of interacting systems of dislocations. Today, experimental microscopy methods can reconstruct three-dimensional images of complex dislocation networks. Concise methods for further analysis and quantitative characterization of those networks, however, are still lagging behind. This makes the linkage to simulation methods difficult.

Nowadays, a number of well-established simulation methods are able to take the dislocation microstructure (directly or indirectly) into account. As a standard method on the nano-scale molecular dynamics (MD) predicts the trajectory of atoms which interact through forces due to a given potential function. Dislocation lines can be reconstructed in a post-processing step, 
which allows MD to reveal great details of dislocation microstructures (e.g., \citep{Prakash2015_ActaMater92}).
The mesoscale, discrete dislocation dynamics (DDD) method does not resolve separate atoms, but represents dislocations as polygons or splines \cite{Lepinoux:1987tj,Ghoniem:1988wu,Devincre:1992iu,Zbib:1998ub,Weygand:2002tq,Po:2014en} that move and interact according to the elastic theory of dislocations \citep{Peach:1950tm}. Both methods, although very powerful and physically detailed, are limited by the number of interacting objects. Furthermore, direct comparison between dislocation microstructures obtained by the two methods, by alternative methods, or with experimental data cannot be done satisfyingly. 

Continuum models of dislocations, on the other hand, do not aim at resolving separate dislocations. Thus, by representing an averaged picture of these linear defects they can operate on larger length scales than MD or DDD. 
Conceptually, one may  roughly separate these models into two classes: local models that are not able to represent fluxes of dislocations,
and non-local continuum methods which govern the flow of dislocations through transport equations \cite{El-Azab2000_PhysRevB61,Arsenlis2004_JMPS52, Reuber2014_ActaMater71,Xiang2009728,Sandfeld2010_JMR, Hochrainer2014_JMPS}. 

While the first class plays an important role in the engineering community, its predictive power is limited since only strongly averaged details of systems of dislocations can be represented, and dynamical aspects due to dislocation motion can not be captured at all. Members of the second class of continuum models are rather designed in a bottom-up approach utilizing (statistical) averaging of discrete dislocations and can, as a consequence, include many significant information about the dynamics of systems of dislocations, i.e., how dislocations move and interact. These ``continuum dislocation dynamics'' (CDD) models are able to simulate arbitrary dislocation densities and large time spans, and make good candidates for complementing MD or DDD simulations.

Direct validation of continuum methods  on the level of dislocation microstructure could up to now only be done for simplified two-dimensional DDD simulations with point-like edge dislocations \citep{Yefimov2004_JMPS52,Hirschberger2011_MSMSE19,Sandfeld2013_MSMSE}. Comparisons with experimental data have been done based on EBSD measurements from  which ``geometrically necessary dislocation'' (GND) densities can be inferred \citep{Ruggles2016_IJP76}. 
Systematic comparisons and validations of continuum methods with lower scale methods or experiments that take all, possibly curved, dislocations into account and consider spatial details of the microstructure were not possible until recently. One of the reasons is that no appropriate continuum description of systems of curved dislocations existed before. 
Using data from lower scale methods or experiments either as initial values, for verification purposes, or, via data mining, as parameters for other simulations, however, is a highly desirable goal. To achieve this two prerequisites must be met:
\begin{enumerate}
	\item[(i)] A \emph{methodology} consistent with the underlying physics for systematic data extraction from lower scale models and/or experiments is required.
	\item[(ii)] A \emph{data format} is needed that is ``rich'' enough to represent even complex microstructures and that allows for detailed analysis and direct comparisons. 
\end{enumerate} 
In principle, one could use dislocation data from DDD or possibly even from CDD models for determining positions of atoms (a dislocation is the boundary of a slipped area within which atoms are displaced by the size of the Burgers vector $\Bb$).
Comparing dislocation microstructures on the level of single atoms, though, is neither practical nor reasonable since in general it is not clear how to define when atomic structures are ``similar'' (e.g., what is a ``large'' deviation? Should it be measured in number of atoms that are not in place or is it the cumulative misplacement of single atoms?).
The main problem here is rather a conceptual problem: although dislocations consist of regions of displaced atoms, they are objects whose main characteristics become visible on the ``mesoscale'', i.e., the scale intermediate between the scale of single atoms and the scale where concepts of continuum mechanics are applicable.
On this scale, dislocations need to be treated as mathematical lines together with the respective energy density fields as, e.g., in DDD simulations.
On this scale, the \emph{collective behavior} is responsible for many emergent properties as, e.g., size effects, hardening or dislocation pattern formation. 

Therefore, we propose a different approach towards characterizing, validating, and data mining of dislocation data based on one of the information-richest CDD models that resulted from the theory by Hochrainer \citep{Hochrainer2009_ICNAAM, Sandfeld2010_JMR, Hochrainer2014_JMPS, Hochrainer2015_PhilMag}. Although this CDD model is able to describe the evolution of dislocations in great detail \citep{Sandfeld2015a_MSMSE23}, we only will use the underlying field variables consisting of density and line curvature data (and possibly higher-order moments thereof). Together with converting data from lower scale methods or experiments into these mesoscale field variables we arrive at a description that is -- unlike discrete dislocation data -- defined in each point of the volume under consideration, which, e.g., allows to simply ``take the difference'' between two data sets. The variables of this CDD model are additionally well suited for statistical averaging and representing ensemble averages, which is an important prerequisite for systematic data mining. For converting properties of discrete lines to continuous field data we use the recently introduced \emph{discrete-to-continuum} (\DtoC) methodology \citep{Sandfeld2015a_MSMSE23}. Among the unique properties of CDD is the hierarchical structure of the theory, where based on the respective requirements different numbers of field variables can be included. This allows for a tailored information content which we view as a physically based approach towards compressing dislocation data.

In the following we first introduce the relevant field variables of the CDD theory, followed by a brief summary of the \DtoC{} methodology for converting discrete dislocations into a continuous field description. Subsequently, the theoretical concepts are applied to a number of numerical examples and benchmark test that show how dislocation configurations of different complexity can be represented and analyzed. We conclude with a detailed discussion and outlook.

\section*{METHODS}

\subsection*{Field Variables of the \CDD{} Theory}
Hochrainer's CDD theory was derived based on statistical averaging of systems of discrete dislocations. In the original, higher-dimensional theory (\hdCDD) the resulting continuous field quantities were two density-like field variables \cite{Hochrainer2007_PhilMag,Sandfeld2010_PhilMag90,Sandfeld2015_IJP}: the total density $\rho\rphi$ and curvature density $q\rphi$ (or alternatively: mean dislocation curvature $k\rphi$), defined in a higher-dimensional space, which, besides the spatial points $\Br$, also contains the line orientation $\varphi$ as extra-dimension. This makes the important information about an average line direction $\Bl(\vp):=[\cos \vp(\Br), \sin \vp(\Br)]$ available. Due to the high computational cost of this detailed formulation most numerical computations were only possible in ``2.5D'' configurations \cite{Sandfeld2010_PhilMag90,Sandfeld2015_IJP, Sandfeld2015_MRS}. To overcome this limitation, a number of so-called simplified \CDD{} formulations were derived, where a reduced amount of information is contained in additional field variables \cite{Hochrainer2009_ICNAAM,Sandfeld2010_JMR,Hochrainer2014_JMPS,Monavari2014_MatResSoc}.

Every \CDD{} theory variant also contains the corresponding evolution equations for the field variables. However, in this work the goal is to use only the field variables for representing dislocation microstructure as static configurations in time. Therefore, the evolution equations will not be introduced here. A convenient property of \CDD{} is that formulations with different degrees of information content can be derived within a hierarchical framework of equations, either based on Fourier series of $\rho\rphi{}$ and $q\rphi{}$ \citep{Hochrainer2014_JMPS} or based on alignment tensors \citep{Hochrainer2015_PhilMag, Monavari2016_JMPS}. In the following we introduce the first order \CDDI{} in more detail and note how higher order theories may be constructed.

\emph{The first order theory \CDDI{}} contains three variables \citep{Hochrainer2014_JMPS,Hochrainer2015_PhilMag}: the total density $\rhot \equiv \rhoZero$, the first order dislocation density alignment tensor $\rhoOne$  and the curvature density $\qt \equiv \qZero$.
Without loss of generality we assume that the (local) coordinate system is aligned with the edge and screw directions. Then, the alignment tensor consists of the signed edge and screw excess (geometrically necessary) dislocation densities, $\rhoOne=[\rho^{\rm e}, \rho^{\rm s}]$. The variables can be obtained from \hdCDD{} as
\begin{align}
	\label{eq:rhot}
	\rhoZero(\Br)&=\int_{0}^{2\pi} \,\!\! \rho\rphi \,\text{d}\vp \\
	\label{eq:rho1}
	\rhoOne(\Br)        &=\int_{0}^{2\pi} \,\!\!\rho\rphi\Bl(\vp) \,\text{d}\vp \\
	\label{eq:qt}
	\qZero(\Br)  &=\int_{0}^{2\pi} \,\!\! q\rphi \,\text{d}\vp
\end{align}
From \eqref{eq:rho1} the classical Kr\"oner-Nye dislocation density tensor can be obtained as $\Balpha = \rhoOne \otimes \Bb$ (where $\otimes$ is the tensor product), the total GND density is $\rhoG=|\rhoOne|$. 
The line curvature can be obtained from $k=q^{(0)}/\rho^{(0)}$.

\emph{Higher order theories \CDDn{}} include fields with additional information, e.g., about orientations of dipole and anisotropic dislocation configurations.
They can be obtained in a systematic manner using
\begin{align}
    \Brho^{(n)}(\Br) &= \int_{0}^{2\pi} \rho\rphi {\Bl(\vp)}^{ \otimes n}
    \,\mathrm{d}\vp,
    \label{eq:An} \\
    \Bq^{(n)}(\Br) &=\int_{0}^{2\pi} q\rphi [\Bve \cdot \Bl(\vp) \otimes \Bl(\vp)^{ \otimes n-1}]^{\mathrm{s}}\,\mathrm{d}\vp,
    \label{eq:qn}
\end{align}
where $(\bullet)^{\otimes m}$ denotes the $m$-times tensor product of $(\bullet)$, the superscript $^\mathrm{s}$ denotes the symmetric part of a tensor, ``$\cdot$'' is the contraction of two tensors, $\Bve$ is the second order Levi-Civita tensor in the slip plane coordinate system, with components $\varepsilon_{ij} = \varepsilon_{ijk}n_k$, and the $n_k$ are the components of the slip system normal $\Bn$, where the Einstein summation convention applies. 
For further details please refer to \citep{Hochrainer2015_PhilMag, Monavari2016_JMPS}.

\subsection*{The Discrete-to-Continuum (\DtoC) Method}
For obtaining the above introduced field variables from discrete simulation methods or experiments one may assume that dislocations have already been identified and their data is available in form of parameterized lines. Following the strategy introduced by one of the authors in \cite{Sandfeld2015a_MSMSE23}, CDD fields can be obtained by 
\begin{enumerate}
	\item[(i)] discretizing the domain into voxels of volume $\Delta V$
	\item[(ii)] computing the line length $L_i$, the average line orientation $\Bxi_i$ and the average curvature $k_i$ for each line segment $i$ within each voxel
	\item[(iii)] computing  CDD quantities for each segment $i$, e.g., $\rhoZero_i=L_i / \Delta V$, $(\rhoOne)_i=\rhoZero_i \Bxi_i$ and $\qZero_i=\rhoZero_i k_i$
	\item[(iv)] integrating or averaging over all $N$ line segments within a voxel, e.g., $\rhoOne=\sum_{i=1}^N (\rhoOne)_i$. 
\end{enumerate}
If the data is intended for use in simulations this may be followed by a convolution of the whole domain with, e.g., a Gaussian function for numerically smoothing the voxel data such that numerical derivatives can be easily obtained.

\section*{RESULTS}
To explain the information content of the \CDDI{} fields we will now set up three quasi-2D test configurations, which represent thin slices of a volume.
This is then followed by a realistic 3D example where the benefit of ensemble averaging is explained.

\subsection*{Basic Test Configurations}
\begin{figure*}
    \centering
    \includegraphics[width=0.95\textwidth]{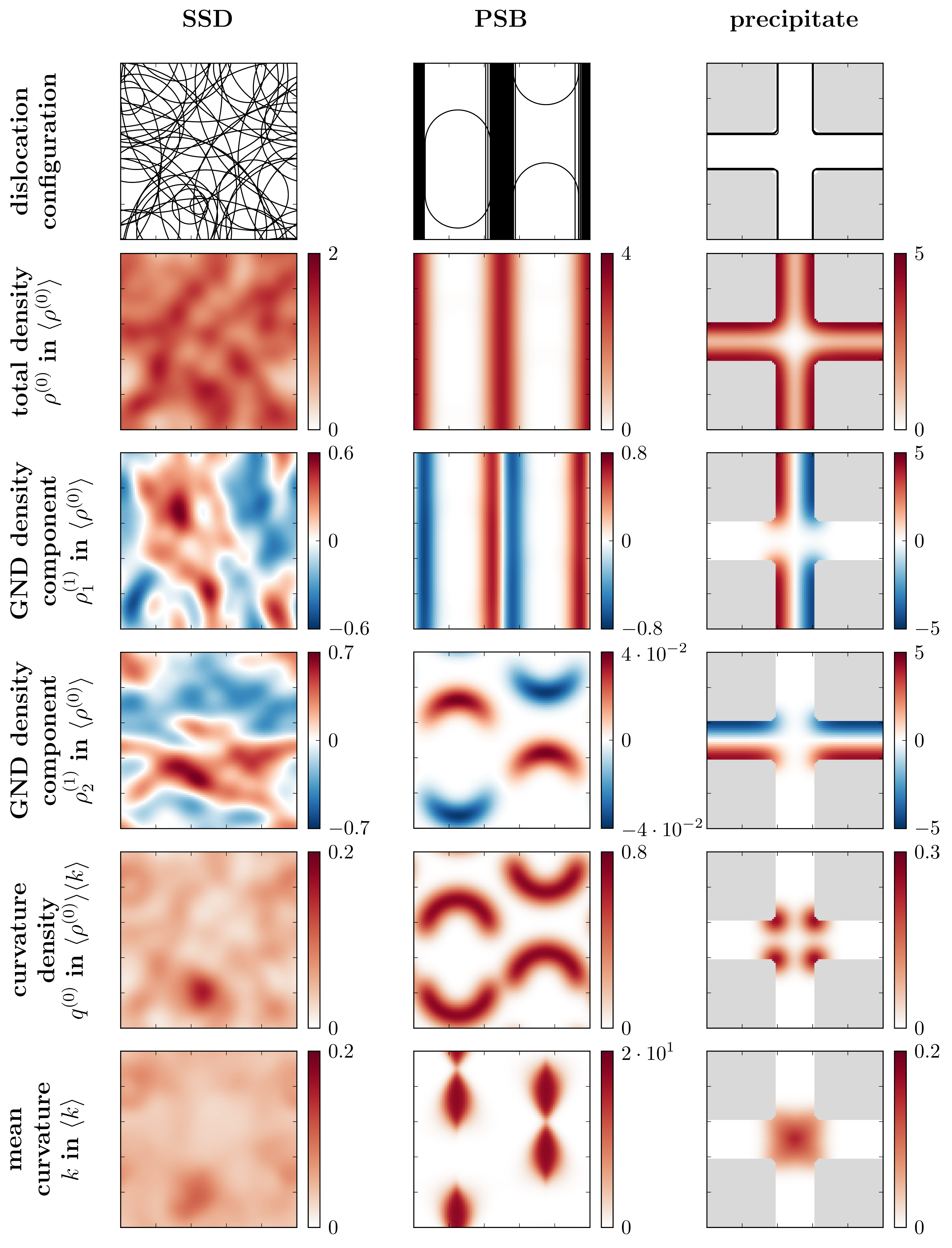}
    \caption{
        Field variables of the first level of the \CDD{} theory, \CDDI{}, for different periodic dislocation structures. 
        Density measures are scaled with the mean of the total density, the curvature density is scaled with the mean of the total density and the mean curvature of the system, and the curvature is scaled with the mean curvature of the system. The gray area in the right column represents the precipitate.}
    \label{fig:simple_configurations}
\end{figure*}

We apply the \DtoC{} framework to three dislocation configurations, exhibiting a number of key features: a distribution of statistically stored dislocation (SSD) loops, wall and channel dislocations within an idealized persistent slip band (PSB) and an idealization of dislocations in a $\gamma/\gamma'$ superalloy.
All systems exhibit periodic boundary conditions and the continuum fields are regularized using a Gaussian function with a standard deviation of 5\,\% of the system size.
The dislocation configuration of these systems as well as the resulting fields used by the \CDDI{} theory are shown in \figref{fig:simple_configurations}.
In the following the generation of the structures are outlined and relevant features of the CDD fields are discussed.

\emph{The SSD configuration} consists of a random distribution of 20 circular dislocation loops with radii ranging from 1/50 to 4/5 of the system size within the same glide system. Varying radii were chosen to demonstrate the resulting fluctuations of the curvature density and the curvature in otherwise relatively homogeneous systems.
The system exhibits spatial fluctuations in the total density as well as in the GND density. There, the distinction between SSDs and GNDs depends on the chosen resolution:
if the size of the averaging voxels equals the size of the whole system, the GND density will be zero because opposite loop segments cancel each other out perfectly.
A smaller averaging size, on the other hand, reveals more local features, which shows in fluctuation of the density fields. Here, the actual resolution is given by the width of the regularization function (for an example of how the regularization width influences the details of the microstructure see \cite{Steinberger_2016_a}).
Curvature density and mean curvature show a distinct maximum as well as fluctuations. The maximum is due to the combination of a high curvature with a higher than average dislocation density. Curvature fluctuations are due to the fact that loops have different radii, and the curvature density is additionally weighted with the respective total density inside each averaging volume. Therefore, it may not be inferred that a high curvature density generally also results in a high mean curvature at the exact same site. 


\emph{An idealized PSB structure} was constructed by randomly placing 30 straight edge dislocations with positive and 30 with negative line direction into a wall-like structure.
The mean $x$ positions for the two dislocation characters inside one wall was shifted to result in a mean distance of 3\,\% of the system size to capture the dipolar character of the dislocation wall within PSBs.
``Pill'' shaped dislocations were placed between the walls to represent threading screw dislocations.

\begin{figure}
    \centering
    \includegraphics{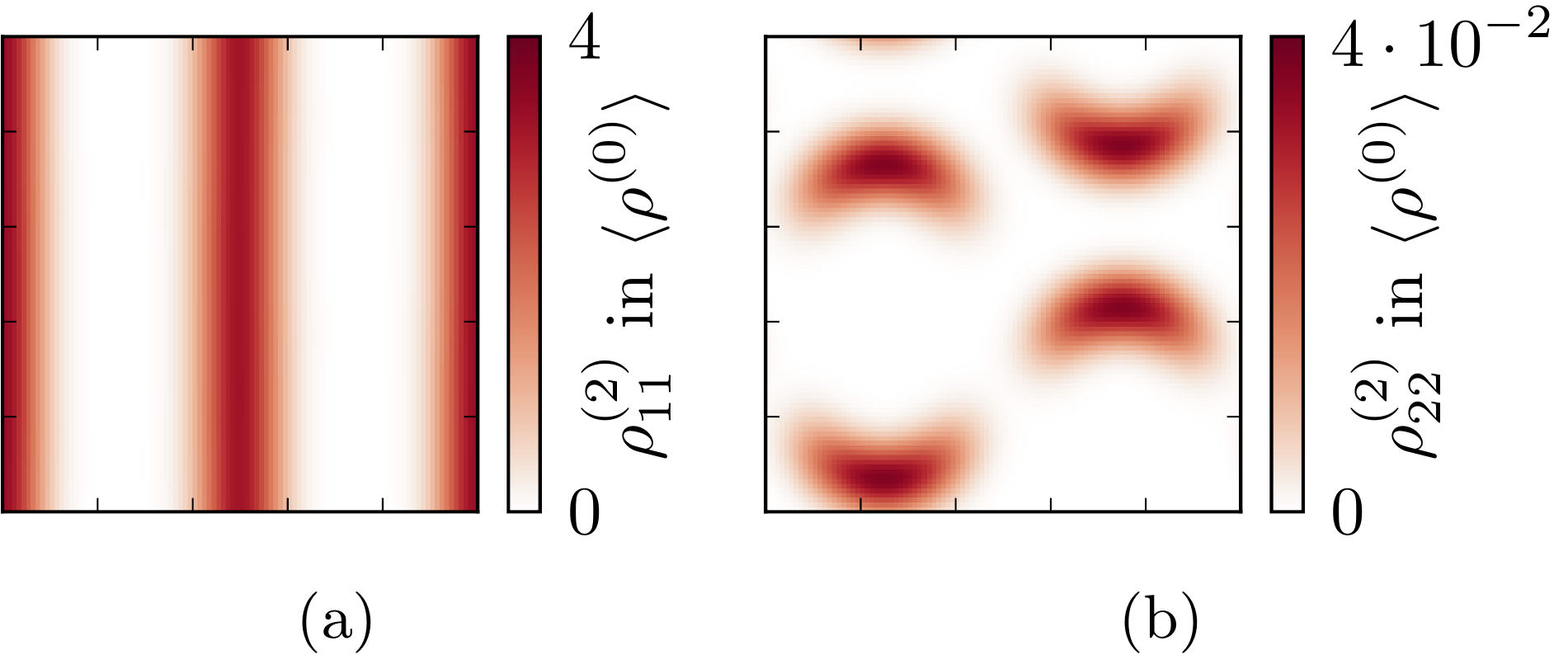}
    \caption{Total (a) edge density $\rho^{(2)}_{11}$ and (b) screw density $\rho^{(2)}_{22}$ for the PSB system.}
    \label{fig:cdd2_psb}
\end{figure}

The system has a large total dislocation density, which is mainly located inside the dislocation wall structure, while by comparison, the contribution of threading screw dislocations is relatively small.
The edge component of the GND density $\rho^{(1)}_{1}$ reveals the polarized character of the dislocation walls with a ``neutral'' axis in the middle of the dislocation wall.
The screw component of the GND density $\rho^{(1)}_{2}$ between the dislocation walls is smaller than the edge component by two orders of magnitude and outlines the position of the quasi-discrete threading screw dislocations.
Both the structure of the total as well as the GND density are within the expectation of physical intuition.
As the threading screw dislocations are the only non-straight segments, their shape is reflected in the curvature density of the system.
Contrary to the SSD system, mean curvature and curvature density look quite differently.
The curvature exhibits a lens-like shape in the middle of the channels with a symmetry axis parallel to the wall structure.
This can be explained based on the influence of the straight segments with zero curvature that form the dislocation walls: due to the regularization width used during the \DtoC{} procedure they also contribute to the mean curvature outside of the dislocation wall and decrease the mean curvature in the channels close to the dislocation walls.
As this influence decreases the high curvature of the threading dislocations becomes the sole contributing factor to the mean curvature and a maximum is reached.
This effect is caused by the specific choice of parameters. 
Moreover, the PSB structure may also be used to outline one of the shortcomings of the \CDDI{} theory: Within \CDDI{} fields the exact composition of SSD densities is unknown, i.e., the theory can not differentiate whether the SSDs of the dislocation walls consists of statistically stored edge or of statistically stored screw dislocations.
The next higher order theory, \CDDII{}, \emph{is} able to capture this due to the second order alignment tensor \cite{Monavari2016_JMPS}
\begin{equation}
    \label{eq:rho2}
    \rhoTwo(\Br) = \int_{0}^{2\pi} \rho\rphi\Bl(\vp) \otimes \Bl(\vp) \,\text{d}\vp,
\end{equation}
whose components $\RHO{2}{11}$ and $\RHO{2}{22}$ correspond to total edge and screw dislocation density, respectively. 
For the PSB system these density measures are shown in \figref{fig:cdd2_psb}. Comparing them to $\rhoZero$ and $\rhoOne$ it can be inferred that the complete dislocation wall comprises edge dislocations, which in the inside of the wall form dipole configurations and further away form polarized edge GND configurations. 

\emph{An idealization of dislocation structure within a $\gamma/\gamma'$ superalloy} is represented by a square precipitate which is impenetrable for dislocations, surrounded by $\gamma$ channels in which dislocation are located. The width of the channels amounts to 20\,\% of the system size, and four dislocations of the same line character are placed in the channels around the precipitate.
It can be seen that the total dislocation density is concentrated at the interface between matrix and precipitate.
The edge component of the GND density is high at the vertical interface (i.e., perpendicular to the Burgers vector), and the density of the screw GND component is concentrated at the horizontal interfaces.
The curvature density exhibits extrema at the corners of the precipitate and is zero at the straight edges further away from the corners.
It may, however, be counter intuitive that the mean curvature exhibits a maximum in the center of the crossing of the channels, while the curvature density is largest close to the corners of the precipitate.
The reason for this is the same as the one given for the PSB system.
Due to the regularization width straight segments still contribute to the mean curvature at the precipitate corners.
Their contribution, however, is negligible in the center of the crossing -- the point furthest away from the straight dislocation sections, which is still fairly close to the corners.

\subsection*{Three-Dimensional Configurations}
The test cases introduced before are useful for elucidating which geometrical properties of systems of dislocations can be described by CDD.
We now turn to a more realistic, three-dimensional system, where dislocations evolve and interact, and we show how \DtoC{} can be used together with ensemble averaging to  differentiate between emergent and random dislocation structures.

\begin{figure*}
    \centering
    \includegraphics{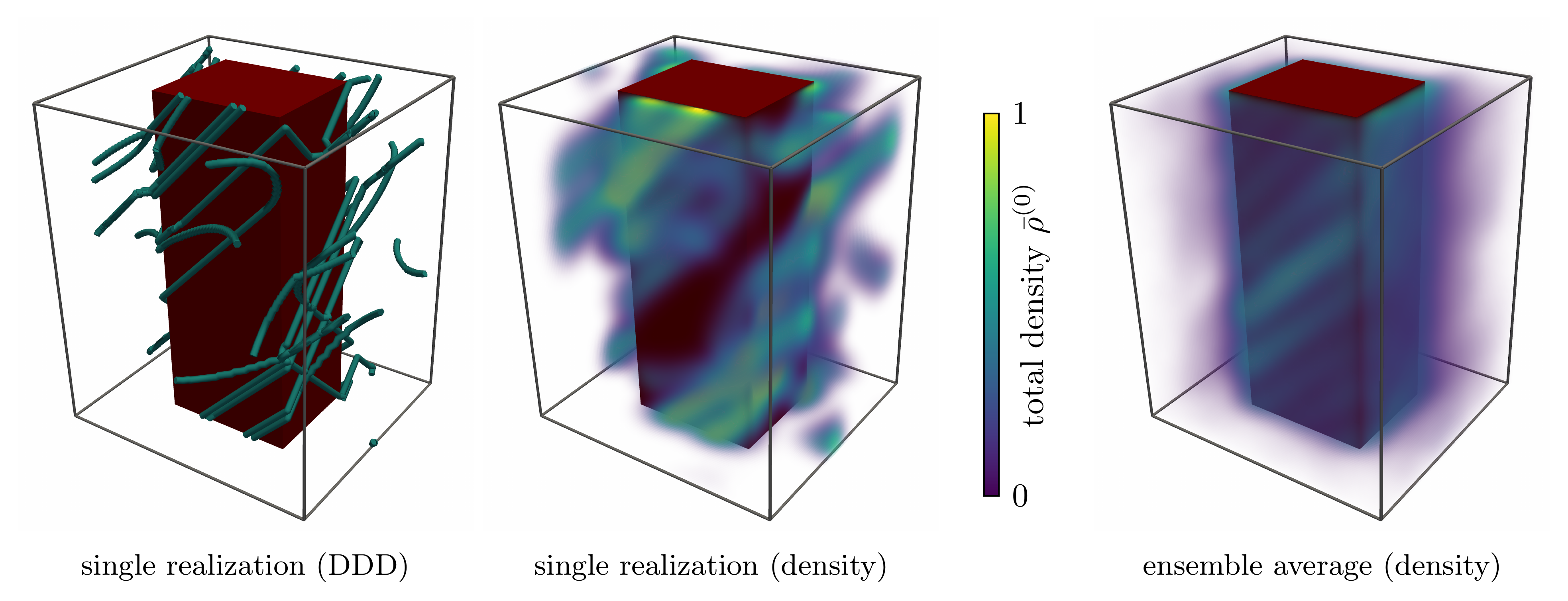}
    \caption{The \DtoC{}-based ensemble averaging of statistically equivalent DDD simulations reveals intrinsic structure. The precipitate is shown in red.}
    \label{fig:ensemble_average_showcase}
\end{figure*}

The system is a periodic 3D simulation box with a size of $17717 \times 19550 \times 23445$ Burger's vector lengths.
It contains an impenetrable, pillar-like precipitate along the $z$-axis.
As initial values for each simulation we place two dipolar loops (consisting only of dislocation segments with edge character) on different slip systems at random positions.
The simulation was evolved with a constant shear strain rate of $20\,\mathrm{s}^{-1}$ using the microMegas DDD code \cite{Devincre_2011_a}.
Material parameters were taken from copper.
After a simulated time of $10\,\mathrm{\mu s}$ the simulation was stopped and the final configuration converted into continuum fields on a $38 \times 42 \times 50$ mesh.
The dislocation structure and the corresponding dislocation density of an example system, and the density of an averaged system of 133 simulations are shown in \figref{fig:ensemble_average_showcase}.

\begin{figure*}
	\centering
	\includegraphics{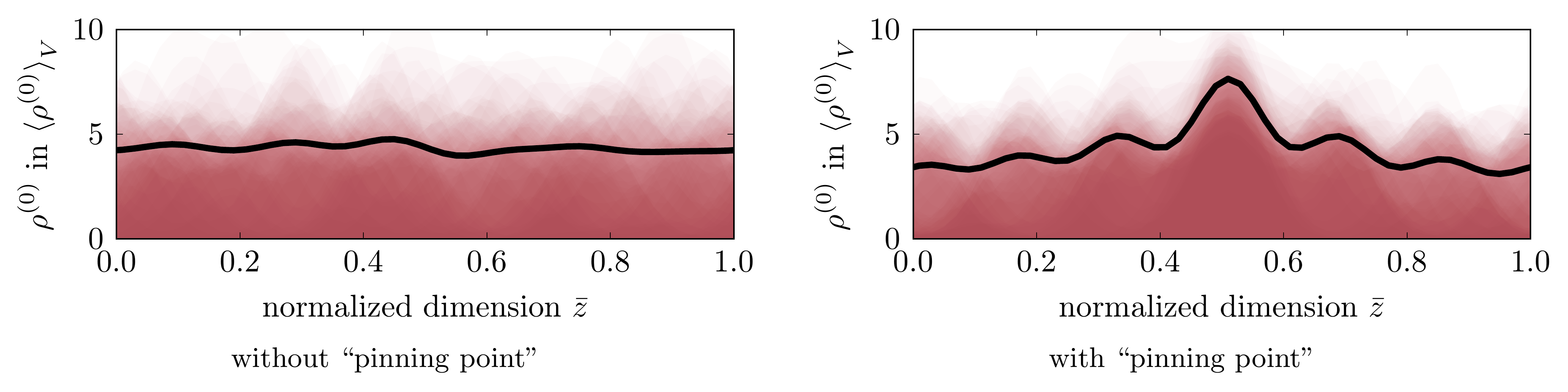}
	\caption{Line profiles of  total dislocation density (red) as well as their ensemble average (black) for 60 realizations. The normalized direction $\bar{z}$ runs along the precipitate axis close to the matrix-precipitate interface. The left and right plots are without and with shift of the maxima. Values are normalized with the mean total dislocation density of the respective 3D system, $\langle \rho^{(0)} \rangle_{V}$.}
	\label{fig:profiles}
\end{figure*}

It can be seen that dislocations are deposited at the interface of the precipitate and also form somewhat regular structures in the matrix.
Along the precipitate interface pair-like dislocation arrangements, single dislocations and a dislocation free zone can be observed.
The continuum representation of this system exhibits very similar features.
From just this one system it might be assumed that the characteristics of this ensemble are dislocations, which  form structures across the matrix and dislocations, being irregularly arranged close to the precipitate.
To investigate this, 133 simulations are conducted and ensemble averages of the respective CDD fields were computed.
Due to periodic boundary conditions special care has to be taken when averaging the systems.
If the ensemble average is computed without taking into account the fact that the system is translationally invariant in the direction of the pillar due to periodic boundary conditions, the result is a rather homogeneous density, as can be seen in the left part of \figref{fig:profiles}.
Using an artifical ``pinning point'' by periodically shifting the dislocation densities along the precipitate until the maxima align, a regular dislocation arrangement at the matrix-precipitate interface can be revealed.
In the near vicinity of this ``pinning point'' a clear maximum can be seen followed by a regular sequence of low and high density with decreasing amplitude.
This can be nicely seen in the right part of \figref{fig:profiles}.
This non-uniform dislocation density distribution at the matrix-precipitate interface is hence characteristic for the strain rate and for a total simulation time of $10\,\mathrm{\mu s}$ we considered (for larger simulation times a regular distribution of dislocations at the interface is expected, but with shorter wavelength and smaller amplitude).
In the matrix itself the density is low compared to the region in the vicinity of the precipitate.
Thus, it follows that the irregular arrangements of the dislocations around the precipitate seen in \figref{fig:ensemble_average_showcase} are not characteristic for this ensemble; dislocations crossing the channel do not form regular structures but are arranged randomly.

\section*{DISCUSSION AND CONCLUSION}
With simulations becoming more realistic and getting closer to experiments the complexity of required initial conditions increases. If these are constructed in a wrong manner (e.g., not consistent with the underlying theory) simulation results might be completely wrong or even the numerical scheme might fail. The \DtoC{} framework is a way for converting discrete dislocation structures obtained from experiments or DDD simulations, which guarantees physically correct initial values. We anticipate that this and similar strategies will become indispensable for simulations in the future.

In the previous section we showed that the information contained in \CDDI{} is already high enough to properly characterize many relevant physical properties of given, idealized microstructures, such that they can be, e.g, used as initial values. But we also found, that there may be situations where the information content of \CDDI{} is not sufficient and we included extra information only contained in the next higher order theory, \CDDII{}. Including more and more information, say from \CDDn{}, we are able to approximate the real, discrete microstructure in all details -- which is nice but of course computationally undesirable. Reversing this direction, i.e., starting from a (virtual) \CDDn{} and decreasing $n$ one can also think of this as a type of data compression. Of course one looses information, but for each order reduction $n\mapsto n-1$ the amount of data that needs to be stored is reduced, e.g., while \CDDII{} requires 7 scalar fields to be stored in the computer ($\rhoZero$, $\RHO{1}{1}$, $\RHO{1}{2}$, $\RHO{2}{11}$, $\RHO{2}{12}$, $\RHO{2}{22}$ and $\qZero$), the amount of data required for \CDDI{} is reduced by $\approx40\,\%$ ($\rhoZero$, $\RHO{1}{1}$, $\RHO{1}{2}$,  and $\qZero$). Since one has complete control over which details are lost, we call this order reduction a ``physical'' compression of dislocation microstructure data. Note, that so far only $\qZero$ (and no higher order terms of $q$) was used as independent field variable within the evolution equations of \CDDI{} and \CDDII{}. Since \DtoC{} only uses the field variables and not the evolution equations of CDD, it might be interesting to study which microstructural details can be included by additionally considering, e.g., $\qOne$.

Sandfeld and Po showed in \cite{Sandfeld2015a_MSMSE23} how the \DtoC{} strategy can  be used to directly validate CDD evolution equations: running a DDD simulation alongside with a CDD simulation (with initial values from \DtoC) the former can be taken as a reference solution, where converting multiple time steps of the discrete system to the corresponding continuum fields via \DtoC{} allows a quantitative comparison with the system evolved by the CDD framework. This is particularly useful in situations where no analytical solution is available or where complex behavior might lead to counter intuitive results. \DtoC{} is a useful tool for the validation of a CDD model and its implementation.

Dislocation-based plasticity in the micro meter regime is strongly interconnected with size-dependent and intermittent behavior together with scatter of, e.g., the flow curves during plastic deformation.
Responsible is the heterogeneous dislocation microstructure.
In general, the spatial or temporal fluctuations can not be seen in continuum simulations, since  a density-based continuum model can be envisaged as representing an ensemble average of statistically equivalent simulations. 
Our 3D configuration nicely shows that \DtoC{} is also useful for computing and analyzing ensemble averaging, when one tries to decide whether a specific realization shows behavior that is or is not characteristic or typical for the ensemble.
Furthermore, in many situations an ensemble average might already be sufficient such that one does not need to store all data from all separate simulations, which might not be admissible due to limited storage memory.
For example, the file holding all snapshots in time of the single system shown in \figref{fig:ensemble_average_showcase}, which is a system with very few dislocations, requires 11\,MB of hard disk space.
More complex systems result even in files of several GB in size. A \CDDI{} file is roughly around 10\,MB, but does not increase regardless how many dislocations are contained.
Additionally, averages over many simulations can be done successively: instead of storing discrete data of all systems and time steps before averaging them, only the average of each time step is stored and ``improved'' with each additional discrete simulation.

Our vision is to extend the interfaces of the \DtoC{} framework such that it can easily read and process data from most common open source DDD and MD codes -- possibly even 3D TEM dislocation data. Furthermore, we plan to extend the capabilities of \DtoC{} towards computation of dislocation  stress fields within arbitrarily shaped, finite domains, which will  add another level of detail to our characterization and  analysis toolbox.
Following the progress is possible through the ``\DtoC'' website \cite{D2C_website}.

\section*{ACKNOWLEDGEMENTS}
Financial support by the German Research Foundation (DFG) through the European M.ERA-NET project FASS, grant no. SA2292/2, is gratefully acknowledged.


\end{document}